# Misrepair-accumulation theory: a theory for understanding aging, cancer-development, longevity, and adaptation

Jicun Wang-Michelitsch[1]*, Thomas M Michelitsch[2]

[1] Department of Medicine, Addenbrooke's Hospital, University of Cambridge, UK (work address until 2007)

[2] Institut Jean le Rond d'Alembert (Paris 6), CNRS UMR 7190 Paris, France

## Abstract

Misrepair-accumulation theory is a novel theory in interpreting aging phenomena. We will introduce the importance of this theory in the present paper. With a new concept of Misrepair, this theory reaches a unified understanding of aging: aging of an organism is a result of accumulation of Misrepairs of its structure. Misrepair-accumulation theory (MA theory) is distinct and it can improve our understanding of life on several aspects. **I.** MA theory is fundamentally different from other theories by: **A.** proposing a generalized concept of Misrepair; **B.** pointing out the evolutionary advantage of aging mechanism; and **C.** distinguishing between aging of a cell and aging of a tissue. **II.** MA theory overcomes the limitations of traditional theories on interpreting of the phenomena of tissue fibrosis and cancer-development. Tissue fibrosis, as a result of accumulation of Misrepairs with collagen fibers, is a powerful evidence for the central role of Misrepairs in aging. Cancer-development is a result of accumulation of DNA Misrepairs in somatic cells, which needs to proceed over many generations of cells in a regenerable tissue. **III.** MA theory improves also our understanding on longevity, premature aging, adaptation, and species' evolution. The potential of longevity of an organism is hidden in the complexity of its structure. An animal has limited longevity because it has limited structural complexity. Limited structural complexity and limited longevity are essential for the survival of a species. Premature aging is a consequence of mis-construction of tissues/organs during development as a consequence of genetic disorder. The abnormality of tissue structure is the common point between premature aging and normal aging. Misrepair mechanism is essential in maintaining the diversity of genome DNAs in a species, thus it is important for species' adaptation and species' evolution. **IV.** MA theory suggests that for retarding aging, the most important is to reduce the opportunity of damage-exposure.

## Keywords

On aging research, some critical questions are not yet satisfactorily answered. These questions are: how we age, why we cannot avoid aging, and why we have limited longevity. Recently, we proposed a Misrepair-accumulation theory (MA theory) for interpreting aging [1, 2]. With a new concept of Misrepair, this theory is fundamentally different from other aging theories. With Misrepair mechanism, this theory can give good answers to the above questions and can give explanations on many aging-related phenomena, including cancer-development, tissue fibrosis, and premature aging. In the present paper, we will make a brief introduction of our interpretations of aging diseases and longevity by this theory. Our discussion tackles the following subjects:

I. A novel aging theory: Misrepair-accumulation theory (MA theory)

II. Distinct points of MA theory

    2.1 A generalized concept of Misrepair
    2.2 Evolutionary advantage of aging mechanism
    2.3 Distinguishing between aging of a cell and aging of a tissue

III. Interpretations of aging diseases and cancer-development by MA theory

    3.1 Tissue fibrosis and fibrosis-associated aging diseases
    3.2 Atherosclerosis
    3.3 Cancer-development

IV. Interpretations of premature aging, longevity, adaptation, and species' evolution by MA theory

    4.1 Premature aging
    4.2 Limited longevity
    4.3 Individual adaptation, species' adaptation, and species' evolution

V. Strategies to retard aging proposed by MA theory

    5.1 Controversial strategies on retarding aging
    5.2 A most important strategy: reducing the opportunity of damage-exposure

## I.    A novel aging theory: Misrepair-accumulation theory (MA theory)

For interpreting aging, we have proposed a novel aging theory, Misrepair-accumulation theory [1, 2]. In this theory, the term of Misrepair is different from that in "Misrepair of DNA". The new concept of Misrepair is defined as *incorrect reconstruction of an injured living structure.* This generalized concept of Misrepair is applicable to all living structures, including molecules (DNAs), cells, and tissues. Like that in scar formation, Misrepair will take place when a complete repair is impossible to achieve for a severe injury. Misrepair is a strategy of repair for maintaining the structural integrity and increasing the surviving chance of an organism. Without Misrepairs, an individual could not survive to the age of reproduction. Therefore, Misrepair mechanism is essential for the survival of a species.

Misrepair is a repair with altered material and in altered remodeling of a structure. The structural alteration by Misrepair is irreversible and irremovable and thus accumulating. Accumulation of Misrepairs will distort gradually the structure of a molecule, a cell or a tissue, appearing as aging of it. Aging can take place on each level of molecule, cell, and tissue, respectively. However, aging of a multi-cellular organism takes place essentially on tissue level. An irreversible change of the spatial relationship between cells/extracellular matrixes (ECMs) in a tissue is *essential and sufficient* for causing a decline of organ functionality. In summary, aging of an organism is a result of accumulation of Misrepairs on tissue level. Aging of an individual is a sacrifice for species' survival.

## II. Distinct points of MA theory

MA theory is fundamentally different from traditional aging theories, and it has three distinct points: 1). proposing a generalized concept of Misrepair; 2). pointing out the evolutionary advantage of aging mechanism; and 3). distinguishing between aging of a cell and aging of a tissue.

### 2.1 A generalized concept of Misrepair

The term of "Misrepair" is traditionally used to refer to the "Mis-match repair" of DNA, which means that a broken DNA is re-linked by incorrect base-pairs. Misrepair of DNA is a kind of "SOS" repair of DNA, for maintaining the structural integrity of DNA and for preventing death of the cell from failure of DNA. In fact, such kind of "incorrect repairs" exists also in tissues, and scar formation is an example. Scar formation, as an altered remodeling of part of the skin, is a consistent response of our organs to a deep wound. This fact indicates that incorrect repair is a way of repair. Incorrect repair can take place in all types of living structures that are repairable, including DNAs, cells, and tissues. More interesting is that incorrect repair is a common change in aging symptoms, including tissue fibrosis, atherosclerosis, and osteoarthritis. To have a unified understanding of aging changes, it is essential to unify all the different forms of incorrect repairs by a new concept. The term of "Misrepair" is simple, well-known, and understandable; therefore it is suitable for this purpose. A generalized concept of Misrepair is thus defined as *incorrect reconstruction of an injured living structure*. With a clear definition, Misrepair as a process and a result of repair is observable and detectable. We believe that generalization of the concept of Misrepair is a big progress on understanding aging.

### 2.2 Evolutionary advantage of aging mechanism

Aging is always thought to be negative for our life, because it is accompanied with diseases and disability. However, with the concept of Misrepair, we can understand that aging is a sacrifice for survival. Aging is not due to insufficiency of repair, but due to life-saving Misrepairs. Evolutionary theories have earlier assumed that aging of an organism is for group benefit, and saving living resources for reproduction rather than maintaining the body is more beneficial for "group survival" [3, 4]. However, if saving resources is essential for "group survival", a more effective way for "group survival" would be to make all mature individuals die soon after the age of reproduction, by a rapid and programmed death rather than through a long process of aging. In reality, the maximum lifespans of most animals are all 5-10 times longer than the time for development. In fact, survival of a species is based on the survival of individuals. However, without Misrepairs, nobody could survive till reproduction age. Misrepair mechanism is a surviving mechanism for individuals, and it is essential for species' survival. Beneficial for species' survival is the evolutionary advantage of aging mechanism. For human being, without Misrepairs, we could not even survive from a skin wound or a bone fracture. It would be impossible for us to recover from a surgical operation without Misrepairs! Misrepairs increase the security of our life, and we should be grateful to them. Positively thinking of aging! We should not fight aging, since it would only lead to early death. Only when we really understand the mechanism of aging, can we really deal with aging!

## 2.3 Distinguishing between aging of a cell and aging of a tissue

In damage-accumulation theory, aging of an organism is predicted to be a result of aging at cellular and molecular level [5]. Differently, in our view, aging of an organism takes place essentially on tissue level. Aging can take place at the levels of tissue, cell and molecule (DNA), respectively. On tissue level, aging appears as disordering of cells and extracellular matrixes (ECMs); on cellular level, aging appears as deformation of cytoskeleton and change of cell shape; and on DNA, aging appears as accumulation of DNA mutations and alteration of DNA sequence. However, aging of a tissue is not essentially a result of aging of cells or aging of DNAs. An irreversible change of the spatial relationship between cells/ECMs in a tissue is *essential and sufficient* for causing a decline of the functionality of the tissue and the organ. Although remaining of an aged cell can be part of an aged tissue, aging of a tissue does not always require aging of cells and aging of molecules. Aging of an organism is due to aging of tissues. Reduction of tissue functionality by Misrepairs is the pathogenesis of aging-associated diseases. Interestingly, aging of a tissue is often the cause rather than the effect of aging of a cell. There are two reasons for that: **A.** the cells in an older tissue can undergo aging faster because of the reduction of tissue functionality; and **B.** aged cells cannot be easily removed in an old tissue.

## III. Interpretations of aging diseases and cancer-development by MA theory

Why are aging-associated diseases incurable? Our answer is: they are incurable because they are caused by the irreversible structural changes of tissues/organs. With age, accumulation of Misrepairs gradually disorganizes the structures of tissues and leads to reduction of the functionality of organs. Here we take fibrosis, atherosclerosis, and cancer-development as examples to explain how a disease develops by Misrepair mechanism.

## 3.1 Tissue fibrosis and fibrosis-associated aging diseases

Tissue fibrosis is the phenomenon that a tissue has progressive deposition of excessive collagen fibers with age. Tissue fibrosis is a typical aging change, and it is the main cause for some diseases including essential arterial hypertension, benign prostatic hyperplasia and senile chronic bronchitis/emphysema. Fibrosis is known to be a result of repairs of a tissue with collagen fibers. In fact, a repair with collagen fibers for replacing dead cells or injured ECMs is a kind of "Misrepair". Progressive process of fibrosis in an organism demonstrates that: A. a process of Misrepair exists; B. Misrepairs are unavoidable; and C. Misrepairs accumulate. Thus, the phenomenon of fibrosis is important evidence for the central role of Misrepair in aging [6]. In arteriosclerotic hypertension and in chronic bronchitis/emphysema, the deposed collagen fibers in arterial walls and airway walls are used for replacing the broken elastic fibers and myofibers, which are injured during repeated deformations of the walls. Stiffness of arterial walls promotes enlargement of myofibers, leading to the thickness of the walls. In Benign prostatic hyperplasia, the deposed collagen fibers in muscular tissues are used for replacing broken myofibers, which are injured during ejaculations. Muscular fibrosis results in accumulation of prostatic fluid and expansion of the prostate. Wrinkle formation is part of the fibrosis of the skin, and it develops by accumulation of collagen fibers of different lengths. Senile hair-loss and senile hair-whitening are also related to dermal fibrosis, because dermal fibrosis affects the blood supply to hair matrixes.

### 3.2 Atherosclerosis

Atherosclerosis is a disease characterized by the development of atherosclerotic plaques in arterial endothelium. Misrepair mechanism plays an important role in the development of plaques [7]. **I.** Development of an atherosclerotic plaque is a result of repair of injured endothelium. Because of infusion and deposition of lipids beneath endothelial cells, the repair has to be achieved by altered remodeling of local endothelium. Such a repair is a manifestation of Misrepair. During repair, smooth muscular cells are clustered and collagen fibers are produced to the lesion of endothelium for reconstructing an anchoring structure for endothelial cells and for forming a barrier to isolate the lipids. **II.** Altered remodeling (Misrepair) makes the local part of endothelium have increased damage-sensitivity and reduced repair-efficiency. Thus this part of endothelium will have increased risk for injuries, for lipid-infusion and for Misrepairs. Focal accumulation of Misrepairs and focal deposition of lipids result in the development of a plaque. **III.** By a viscous circle between lipid-infusion and endothelium-Misrepair, growing of a plaque is self-accelerating. Namely, once a plaque develops, it grows in an increasing rate with time and it does not stop growing. Within part of the arterial wall, older plaques grow faster than younger ones. Thus old plaques are always bigger than new ones, resulting in an in-homogenous distribution of plaques [8]. The oldest and the biggest plaque is the most threatening one in narrowing local vessel lumen. Therefore the accelerated growing of plaques is a fatal factor in atherosclerosis.

### 3.3 Cancer-development

Development of a tumor is known to be a result of accumulation of DNA changes in somatic cells. However, the processes of how DNA changes are produced and how they accumulate in somatic cells are not clear. DNA changes include two types: chromosome changes and point DNA mutations. However, only point DNA mutations can survive in cells. Thus, they are the main type of DNA changes that can remain and accumulate in cells. Severe DNA injuries are the causes for DNA mutations; however Misrepair of DNA is an essential process for transforming a DNA injury into a "survivable and inheritable" DNA mutation [9]. In somatic cells, Misrepair of DNA is the main source of DNA mutations. Since the surviving chance of a cell from Misrepair of DNA is low, accumulation of DNA mutations can take place only possibly in the cells that can proliferate. Tumors can only develop in the tissues that are regenerable. The accumulation of Misrepairs of DNA needs to proceed over many generations of cells, and cell transformation from a normal cell into a tumor cell is a slow and long process. However, once a cell is transformed, especially when it is malignantly transformed, the deficiency of DNA repair and the rapid cell proliferation will accelerate the accumulation of DNA mutations. The process of accumulation of DNA mutations is actually a process of aging of a genome DNA. Repeated injuries and repeated regenerations of cells are the two preconditions for tumor-development. For cancer prevention, a moderate and flexible living style is advised.

## IV. Interpretations of premature aging, longevity, adaptation, and species' evolution by MA theory

MA theory is a theory not only for understanding aging-associated diseases, but also for better understanding the phenomena of premature aging, longevity, adaptation and species' evolution. Misrepair mechanism is a mechanism that underlies all of these phenomena.

### 4.1 Premature aging

Hutchinson–Gilford Progeria Syndrome, Werner syndrome, and Cockayne syndrome are the three genetic disorders in which the children have premature aging features [10, 11]. Although these syndromes have different genetic backgrounds, the patients all have abnormal structures in tissues and organs like that in normal aging. Therefore, the abnormality of tissue structure is the common point between premature aging and normal aging. Structural abnormality links the defective development and the defective repair (Misrepair) [12]. Defective development is a result of mis-construction of tissues/organs, as a consequence of genetic mutation. Differently, aging is a result of mis-reconstructions, namely the Misrepairs, for maintaining the structure of tissues/organs. Construction-reconstruction of the structure of an organism is thus the coupling point between development and aging. Mis-construction and Mis-reconstruction (Misrepair) are the essential processes in the development of aging-like feathers. Misrepairs are local and only affect the injured part of a tissue. Differently, the mis-construction caused by a genetic disorder is systemic and affects all tissues/organs. The organs that develop defectively in these syndromes have low potential of functionality and will go to failure earlier and faster. A Misrepair is not a "real mistake" but a compromise of an organism necessary for a longer survival. Differently, the mis-construction in development is a "real mistake", and it will lead to failure of development.

## 4.2 Limited longevity

All of the organisms in the world have a limited longevity, and different species' of organisms have different longevities. What is the factor that determines the potential of longevity of a species? What is the factor that determines the individual lifespan? These are the two questions over longevity. Potential of longevity is the maximum longevity of a species if the individuals live in an ideal environment. Longevity of an organism includes the time for development (mature time) and the time for structure-maintenance (maintaining time). The mature time for an organism depends on its structural complexity. The maintaining time for an organism is related to the degree of damage-exposure and the potential of functionality for structure-maintenance. Potential of functionality include the capacity of basic functionality and the potential of functional compensation. The capacity of basic functionality is built in the structural complexity of the organism. However, the structural complexity and the functionality of an organism will be reduced gradually with age by the accumulation of Misrepairs. Functional compensation can slow down the decline of functionality by two strategies: functional substitution and partial regeneration. These two strategies are both built in the structural complexity. The potential of longevity of an organism is therefore hidden in its structural complexity, which is determined by the gene configuration [13]. An animal has limited longevity because it has limited structural complexity. Limited structural complexity and limited longevity are essential for the survival of a species. Despite having the same potential of longevity, individuals of a species have different lifespans. The lifespan of an individual is related to the degree of damage-exposure. Thus, lifespan is determined by the living environment and the living habit of the individual.

## 4.3 Individual adaptation, species' adaptation and species' evolution

As a surviving strategy, Misrepair mechanism plays also an important role in individual adaptation, species' survival, and species' differentiation [14]. **Firstly**, Misrepair of an injury is one of the manners of individual adaptation; and Misrepair mechanism gives an organism a great potential in adapting to changeable and destructive environment. **Secondly**, Misrepair mechanism is important in maintaining and enlarging the diversity of genome DNAs of a species; and a large diversity of genome DNAs is essential for the adaptation and the differentiation of a species in different environments. On one hand, somatic Misrepairs are essential for maintaining the sufficient number of individuals in a species, which are the carriers of different genome DNAs. On the other hand, Misrepair of DNA is a source of DNA mutations and the DNA Misrepairs in germ cells contribute to the diversity of genome DNAs. Therefore, Misrepair mechanism is important in individual adaptation, species' adaptation, and species' evolution.

## V. Strategies to retard aging proposed by MA theory

An important goal to study aging mechanism is to find out a way to delay aging-diseases and extend our lifespans. Several strategies for extending lifespan have been studied on animals; however their effects are controversial. Low-dose aspirin is found to be effective in reducing the risk of vascular diseases and cancer-development; however aspirin has severe side effects. MA theory suggests that for retarding aging, the most important is to reduce the opportunity of damage-exposure.

## 5.1 Controversial strategies on retarding aging

Some strategies have been used on animals for retarding aging, and they include caloric restriction (CR), inhibiting free radicals by antioxidants, and genetic manipulation of aging-related genes. However, the effects of these strategies are controversial. CR was for a long time thought to be a way for extending lifespan. Earlier studies showed that less consummation of energy can extend the lifespans of some animals such as mice, monkeys, rodents and primates. However more and more studies do not support this idea [15, 16]. A problem of this strategy is: the animals that have longer lifespans by caloric restriction are often accompanied with abnormal developments including loss of fertility [17]. A fundamental question over CR is: even if CR is effective on extending lifespan, this effect is not necessarily through retarding aging. Another strategy is to eliminate free radicals by antioxidants. Some studies showed that application of antioxidants can extend the lifespans of some species' such as yeast, *Drosophila,* and roundworms [18]. However, this effect of antioxidants is not confirmed. The question is: even if some antioxidants are effective on extending lifespan, the effect is not necessarily through anti-aging.

Genetic manipulation of aging-related genes is a modern strategy for anti-aging, and such kinds of researches are being done by many scientists in the world. Many aging-related genes and lifespan-related genes have been identified [19, 20]. However, the roles of these genes in aging are not clear. There are several defects in the belief in the aging-controlling genes [21]. **A**. Even if some genes are related to lifespans, they are not necessarily involved in aging process. **B**. Even if some genes are aging-related, they are not necessarily the controlling elements of aging process. **C.** Even if some genes are important in regulating aging process, they cannot control the process of aging completely and independently. **D.** If such controlling genes exist, the distribution of aging changes such as age spots should be independent of the locations of damage, which is not true. **E.** If aging is controlled completely by certain genes, when these genes fail in an individual by DNA mutations, this individual could be immortal. However, so far no immortal case has been observed in the world. In our view, the genes that are important in aging process should be also important in the development and in the repair/maintenance of the structure of an organism. Manipulating of these genes may result in defective development like that in premature aging syndromes [22]. Such a strategy has no future.

## 5.2 A most important strategy: reducing the opportunity of damage-exposure

Misrepair-accumulation theory suggests that aging is for survival of an organism from injuries. Fighting aging will only result in early death from defect of repair. We should try to slow down aging rather than to fight aging. A principle for retarding aging is to slow down the rate of accumulation of Misrepairs. There can be three strategies for that: reducing damage-exposure, partial inhibiting of repair, and improving the efficiency of repair/maintenance. It is clear that partial-inhibiting of repair is risky, because it may cause fatal complications such as bleeding, infection, or organ failure. Aspirin is a medicine with effect of anti-inflammation and inhibiting repair. Some studies have shown the effect of low-dose aspirin on reducing the risk of cancers and vascular diseases. In our view, the effects of low-dose aspirin in these diseases are through reducing the rate of accumulation of Misrepairs by partial-inhibiting of repair [7]. The second strategy is to improve the efficiency of repair/maintenance system of cells and tissues by genetic manipulation. But as discussed above, genetic modification may result in defective development rather than improved functionality. In fact, no matter how capable a repair system is, Misrepair will anyway take place when an injury is severe. A full repair of a severe injury may need a too long time to achieve. Thus for retarding aging, the most important is to reduce the opportunity of damage-exposure.

A principle for reducing damage-exposure is to avoid unnecessary damage. Some of our living habits are only to overload our organs and cause unnecessary injuries. These habits include smoking, too much eating, long-time sun-bathing and violent sports. Stop of these habits is the first step to avoid unnecessary damage. Secondly, sleep and rest sufficiently. During sleeping, our body is exposed to the least damage and the repairing processes are not disturbed. Thirdly, do not blindly take anti-aging medicines. Nowadays, one can find a variety of medicines that are marked with effects of anti-aging. Some of these medicines may be really effective on controlling some diseases; however most of them can be toxic rather than anti-aging. Taking such medicines will only increase the burden to the organs and enhance aging of them. Chronic inflammations are the causes for accelerated aging of organs and for cancer-development. Repeated injuries are the promoters for chronic inflammations; therefore we should avoid repeated injuries. For reducing the intrinsic injuries, a moderate lifestyle to avoid overloading the organs is important. For reducing the external injuries, a flexible lifestyle to avoid repeated contacting with the same types of aggressive substances in food and in air is helpful.